# Room-temperature magnetic p-n junctions for charge-and-spin diodes

Yuzhang Jiao[1,2,†], Yutong Wang[3,4,†], Xiangning Du[1,2,†], You Ba[3,4,†], Yingqi Zhang[1], Zhiwei Tang[5], Xiangrong Wang[6], Tiantian Chai[1,2], Xiaoke Mu[5]*, Cheng Song[2]*, Kefu Yao[1], Zhengjun Zhang[2], Yonggang Zhao[3,4]*, Na Chen[1,2]*

[1]Key Laboratory for Advanced Materials Processing Technology (MOE), The State Key Laboratory of New Ceramics and Fine Processing, School of Materials Science and Engineering, Tsinghua University, Beijing 100084, China.
[2]Key laboratory for Advanced Materials (MOE), School of Materials Science and Engineering, Tsinghua University, Beijing 100084, China.
[3]Department of Physics, State Key Laboratory of Low-Dimensional Quantum Physics, Tsinghua University, Beijing 100084, China.
[4]Frontier Science Center for Quantum Information, Tsinghua University, Beijing 100084, China.
[5]School of Materials and Energy and Electron Microscopy Centre, Lanzhou University, Lanzhou 730000, China.
[6]School of Science and Engineering, Chinese University of Hong Kong, Shenzhen, Shenzhen 51817, China.
[†]These authors contributed equally to this work.
*Corresponding authors. Email: chennadm@mail.tsinghua.edu.cn

**Abstract:** Non-magnetic p-n junctions have been fundamental components in the silicon era, serving as the backbone for nearly all Si-based semiconductor devices, including transistors. To tackle challenges such as scaling limitations, excessive latency, and high-power consumption in Si-based electronics, we develop magnetic p-n junctions composed of a p-type amorphous magnetic semiconductor (p-AMS) and n-type Si. These charge-and-spin junctions exhibit typical diode characteristics for charge current, along with distinctive spin diode features. By manipulating spin-polarized space charges, we observed a giant magnetic enhancement of approximately 24.36% at a breakdown current of 5 mA, and an impressive 29-fold increase in magnetic moments for p-AMS. The observed spin behavior is attributed to space charge effects or carrier depletion in the p-AMS with extended hole states.

In addressing the pressing challenges of scaling limitations, excessive latency, and high-power consumption in Si-based semiconductor electronics (*1–4*), it is crucial to explore ultralow-power devices that seamlessly integrate multiple functions, such as information processing, transportation, and storage within a single unit (*5–11*). Analogous to the fundamental importance of non-magnetic p-n junctions in Si-based electronics, the development of magnetic p-n junctions is essential, as they can function as both charge diodes, which switch, rectify, and amplify electrical signals, and spin diodes, performing similar functions for magnetic signals. Magnetic semiconductors hold the promise of merging charge-based and spin-based functionalities, facilitating the development of magnetic p-n junctions capable of rectifying, amplifying, and switching both electrical and magnetic signals (*12–15*).

While existing devices such as spin light-emitting diodes (*16*), magnetic heterojunction diodes (*17*, *18*), spin-Esaki diodes (*19*), and bipolar magnetic junctions (*20*) have demonstrated notable spin-related phenomena, they predominantly focus on magnetic field-dependent transport properties. A significant gap remains in achieving all-electrical control of both electrical and magnetic switching, rectification, and amplification at room temperature in these spin-based p-n diodes (*21–25*).

To overcome this barrier, we introduced high-electronegativity oxygen into high Curie temperature ferromagnetic amorphous alloys (*26*, *27*), resulting in a novel category of amorphous magnetic semiconductors (AMSs). Among these, p-type CoFeTaBO$_x$ AMSs, with oxygen concentration ($x$) ranging from 48.5 to 56.0 at.%, exhibit robust ferromagnetism and elevated Curie temperatures exceeding 600 K. Consequently, we constructed a magnetic p-n junction diode by integrating p-type CoFeTaBO$_{0.485}$ (p-AMS) with n-type Si (n-Si) as shown in Fig. 1A. According to the spin-polarized transport model (*28*, *29*), the spin-splittings (Zeeman or exchange) in p-AMS cause unequal numbers for spin-up and spin-down hole carriers due to a strong interaction between local moments (LMs) and holes (*30*). In contrast, non-magnetic n-Si does not exhibit spin splitting, resulting in electron carriers that are not spin polarized. This configuration forms a magnetic p-n junction via a space charge region, demonstrating the dual functionalities of charge and spin diodes through the manipulation of both electrical current and spin state.

**Magnetic p-n junctions for dual functionalities of both charge and spin diodes**

The p-n junction diode operates as a two-terminal electronic component exhibiting asymmetric conductance, characterized by low resistance in forward bias above a threshold value ($V_t$) of approximately 0.5 V and high resistance in the reverse direction (Fig. 1B). Notably, $V_t$ increases from ~0.5 V to ~1.0 V as the temperature decreases from 300 K to 100 K, reflecting typical behavior of non-magnetic p-n charge diodes (*13*, *14*). A lower $V_t$ at room temperature indicates reduced power dissipation, making this device highly desirable for ultralow-power applications. Thus, this magnetic p-n junction can switch, rectify, and amplify electrical signals, functioning effectively as a conventional charge diode (Fig. 1B).

In addition, the p-n junction diode also functions as a spin diode, exhibiting a spin-based magnetic rectification and switching effect; its magnetization decreases in response to a magnetic field of 50 Oe when the current is activated at $V_\text{t}$, as shown in Fig. 1B. The current-induced magnetization modulation, $|\Delta M|/M_0$, is calculated using the difference between spontaneous magnetization ($M$) with current and the magnetization without current ($M_0$). For all tested temperatures, $|\Delta M|/M_0$ increases with current, peaking at room temperature (Fig. 1B). For instance, at a current of 1 mA, $|\Delta M|/M_0$ decreases from ~6.01% at 300 K to ~3.33% at 100 K. This current corresponds to an ultralow current density of about $2.5 \times 10^{-2}$ A/cm²—6~8 orders of magnitude lower than the requirements for spin-orbit torque or spin transfer torque devices (*9–11*).

To evaluate the stability of room-temperature magnetization switching under a 50 Oe magnetic field, we conducted multiple voltage-controlled electrical switching cycles between 0 and 1 mA (Fig. 1C). As the current toggles, the magnetization transitions from high to low states, achieving a $|\Delta M|/M_0$ value of ~6.01% at 300 K, confirming the unidirectional conduction results. Notably, we found that magnetization switching can be realized solely by current, independent of an external magnetic field as shown in Fig. 1C. As the current shifts from 0 to 1 mA, $M$ decreases, resulting in an approximate $|\Delta M|/M_0$ of 4.24% at 0 Oe. Increasing the magnetic field from 0 to 100 Oe produces similar switching behavior, with a $|\Delta M|/M_0$ value of about 3.27% at 100 Oe and further increases in magnetic field lead to a gradual decrease in $|\Delta M|/M_0$ until saturation is achieved.

Our results indicate that $M$ decreases concurrently with the forward current flowing through the magnetic diode. Thermal effects, typically attributed to Joule heating, could contribute to the reduction in $M$. Based on the *I-U* curve shown in Fig. 1B, a current of 1 mA under an applied voltage of 1.25 V results in a temperature rise of ~$2P_\text{l}/\kappa_\text{sub}$, where $P_\text{l}$ is the thermal power dissipated per unit of length of the thin film, and $\kappa_\text{sub}$ is the thermal conductivity of the substrate (*31*). These calculations suggest that the temperature rise of the p-AMS from a 1 mA current is negligible (approximately 0.0064 K if thermal conductivity of single-crystalline n-Si at room temperature is used). Thus, the reduction of *M* induced by thermal effect is ruled out.

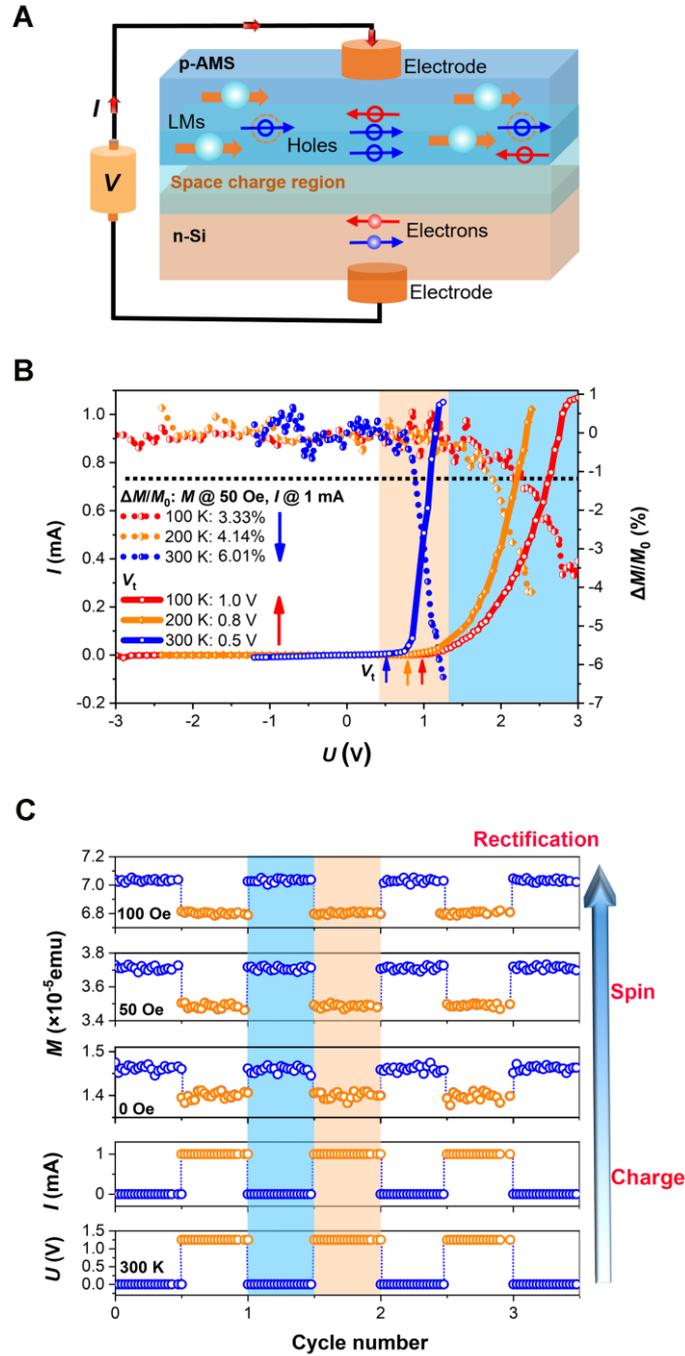

**Fig. 1. Characterization of the magnetic p-n junction diode based on a p-type amorphous magnetic semiconductor (p-AMS) and an n-type single-crystalline Si (n-Si).** (**A**) Schematic diagram of device configuration for the p-n junction diode. (**B**) Current-voltage (*I-U*) curve of the p-n junction diode. The turn-on voltages ($V_t$) are ~0.5, ~0.8 and ~1.0 V at 300 K, 200 K and 100 K, respectively. (**C**) Cycles for room-temperature magnetization (*M*) responses to voltage-controlled current switching between 0 and 1 mA measured at different magnetic fields of 0, 50 Oe and 100 Oe, respectively.

It is reasonable to expect that the reverse current can induce a reverse process of magnetization enhancement in the p-AMS. To investigate this possibility, we measured the *I-U* characteristic curve of the diode over a broader voltage range, revealing a reverse breakdown voltage ($V_{br}$) of approximately 6.5 V, as shown in Fig. 2A. We further explored the dependence of magnetization on the magnitude and direction of the current to uncover the underlying mechanisms of spin-related transport behavior in the diode. To prevent damage to the diode, the reverse breakdown current was limited to a maximum of 5 mA, corresponding to a reverse voltage of about 7 V, as indicated in Fig. 2A.

In comparison to the forward current of 1 mA, we first tested a small reverse current of 1 mA, which required an applied voltage of -4.8 V. As illustrated in Fig. 2B, the magnetization indeed increases from a low state to a high state when the current is switched from 0 to -1 mA, with a $|\Delta M|/M_0$ value calculated to be approximately 4.65%, closely matching the value of 4.24% observed for the forward current of 1 mA. Therefore, this magnetic p-n junction facilitates a complete process for switching and rectification of magnetic signals. Subsequently, we tested the spin-based amplification of magnetization in this magnetic diode.

We utilized a current of 5 mA. Remarkably, even at this reverse breakdown current, the p-n diode operates reliably without any performance degradation. As depicted in Fig. 2B, the switching process proves to be reversible. When the current is toggled from +5 mA to 0, and subsequently to -5 mA within a single measurement cycle, the magnetization transitions from the lowest state to an intermediate state, and then to the highest state. The $|\Delta M|/M_0$ value increases from approximately 7.10% at +5 mA to about 24.36% at -5 mA, demonstrating a significant direction-dependent amplification effect, as illustrated in Figs. 2B and 2C. That is, this magnetic p-n junction realizes dual functionalities of both charge diode and spin diode, accomplishing the property requirements as a charge-and-spin diode.

The magnetization of the p-AMS thin film is influenced by the concentration and spin polarization of the majority holes (*6, 26*). At the same electrical current magnitude, the direction-dependent giant magnetization amplification induced by the breakdown current indicates a significant imbalance in the spin polarization of the majority hole carriers. It is reasonable to infer that the mechanisms generating reverse breakdown current differ from those responsible for forward currents and reverse leakage currents, which contribute to the observed spin-related direction-dependent transport behavior in the charge-and-spin diode. Given the non-magnetic nature of n-Si, we propose that tunnel breakdown occurs through the tunneled valence electrons from the p-AMS, thereby enhancing the spin polarization of charge carriers within the diode. In this context, the space charge region at the magnetic p-n junction could function as a spin modulator, effectively switching, rectifying and amplifying magnetic signals.

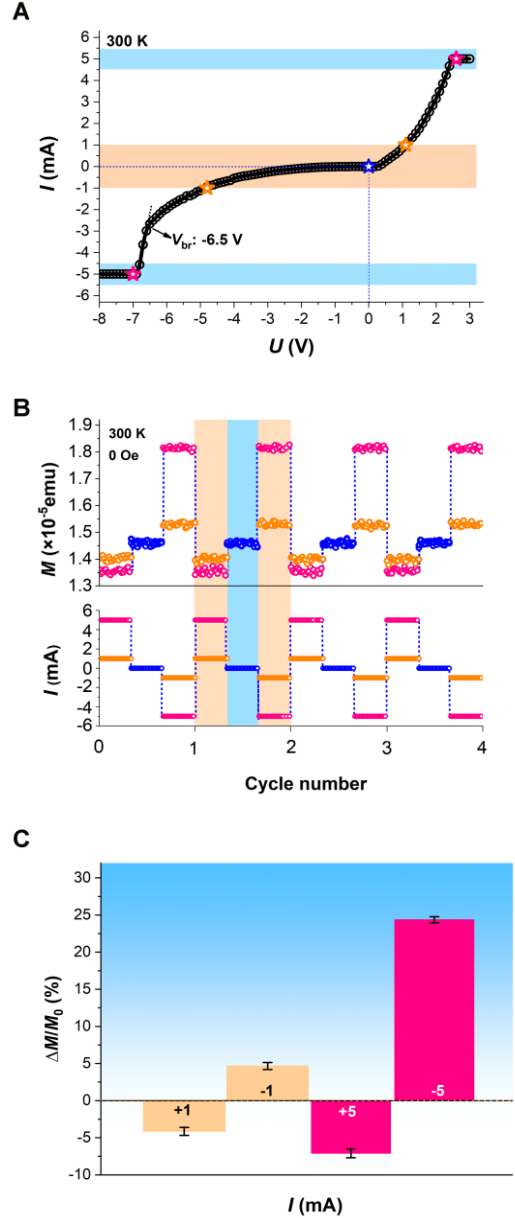

**Fig. 2. Room-temperature direction-dependent spin modulation effect in the magnetic p-n diode under all-electrical control of ferromagnetism without any magnetic field.** (**A**) A complete current-voltage (*I-U*) curve of the magnetic p–n diode characteristic of the reverse breakdown voltage ($V_{br}$) at about -6.5 V. (**B**) Cycles for magnetization (*M*) responses to positive and negative current of 1 and 5 mA, respectively. (**C**) $\Delta M/M_0$ showing significant direction-dependent responses to positive and negative current

## The role of the space charge region as a spin modulator

To further investigate the role of the space charge region in regulating the magnetization states of the diode, we conducted a detailed examination of the p-n junction interface. Figure 3A presents a bright-field scanning transmission electron microscopy (BF-STEM) image of the cross-section at the junction interface, revealing a structural transition from the ordered single-crystalline n-Si to the disordered p-AMS. The homogeneous maze-like atomic arrangements observed in the p-AMS confirm its single-phase amorphous nature, ruling out the presence of a second phase, particularly ferromagnetic nano-precipitates (*32*).

Furthermore, the electric field distribution at the junction interface was mapped, as shown in the differential phase contrast (DPC) image on the left side of Fig. 3B. In DPC imaging, electron beam deflections arise from internal electrostatic potentials within the specimen, allowing the technique to probe local electric fields generated by both intrinsic material potentials and built-in electrostatic fields. The integrated DPC (iDPC) image displays the two-dimensional projected electrostatic potential distribution (see right side of Fig. 3B).

A line profile of the electrostatic potential distribution reveals that the potential on the n-Si side of the space charge region initially decreases to a minimum, resulting in a built-in electric field directed from n-Si toward the interface (Fig. 3C). Upon crossing the interface, the potential begins to rise on the p-AMS side. The emergence of an anomalous minimum in the electrostatic potential distribution causes the electric field strength to transition dramatically from a positive maximum to a negative minimum at the p-n junction. This double-extrema electric field strength distribution induces a triplet-built-in electric field state at the junction, which is distinctly different from observations in conventional non-magnetic p-n junctions (*33*, *34*). The line profile of the charge density distribution map, obtained by calculating the divergence of the electric field map, indicates that negative charges are concentrated at the interface, while weak positive charges accumulate on both sides of the interface. The resulting triplet built-in electric fields create imbalance barriers that impede further diffusion or drift of both electrons and holes. These findings confirm that the formation of the space charge region plays a crucial role in controlling the transport of charge carriers in this charge-and-spin diode.

To further elucidate the underlying physics for the presence of the abnormal minimum potential and negative charge accumulation, we employed first-principles computations to investigate the atomic and electronic structures of the p-AMS. The atomic structural model reveals a disordered amorphous configuration (Fig. 3D). The total density of states (TDOS) indicates the presence of defect bands within the mobility gap (MG), with the Fermi energy ($E_F$) positioned at the top of the delocalized valence band (Fig. 3E). Non-magnetic elements such as Ta and B exhibit higher formation energies for oxides, indicating a stronger affinity for oxygen compared to Co and Fe in the p-AMS. Consequently, the atomic partial density of states (PDOS) plots for both Ta and B show localized valence bands at deep energy levels below the mobility edge. Similarly, the PDOS plots for Fe and O also indicate localized valence bands (Fig. 3F).

As a result, the defect bands within the MG are primarily attributed to the PDOS of the Co 3$d$ bands, which exhibit shallow acceptor states (Fig. 3E). These findings suggest that the $E_F$ level of the p-AMS, situated at the top of the delocalized Co valence bands, is higher than that of n-Si. When integrated, the difference in their $E_F$ levels facilitates minority carrier drift and/or valence electron tunneling from the p-AMS to n-Si. This behavior, which balances the $E_F$ levels at the p-n junction interface, leads to electron accumulation at the junction interface, resulting in the observed minimum electrostatic potential. These results align well with the experimentally obtained mapping results of the electrostatic potential and electric field strength at the space charge region (Fig. 3B and 3C).

The local valence states of the involved elements, particularly the ferromagnetic Co and Fe in the p-AMS, play a crucial role in determining the spin polarization of charge carriers (*35*, *36*). The three-dimensional charge density distribution results indicate that Co exhibits a relatively wide distribution of valence electron transfer, corroborating the extended distribution of defect bands near the valence band maximum (Fig. 3E). To further elucidate the role of the space charge region at the p-n junction, we employed electron energy-loss spectroscopy (EELS) to probe the local valence states of both Fe and Co in the charge-and-spin diode.

The high-angle annular dark field scanning transmission electron microscopy (HAADF-STEM) image of the cross-section reveals a distinct interface region characterized by differences in contrast (Fig. 3G). Three local areas were selected for the collection of EELS spectra (Fig. 3H). Consequently, the average 3$d$-electron occupancy states ($n_{3d}$) of both Fe and Co at the interface region, in areas close to the interface, and in areas farther from the interface were calculated using the white-line ratio method (*37*).

As shown in Fig. 3I, the $n_{3d}$ values of Fe and Co at the junction region differ significantly from those in the regions near and away from the interface. At the interface, the $n_{3d}$ of Co decreases by 0.41 electrons per Co atom, while the $n_{3d}$ of Fe increases by 0.36 electrons per Fe atom, resulting in equal $n_{3d}$ values for Co and Fe. Notably, the concentration of Co in the p-AMS is approximately three times that of Fe. Therefore, the additional transferred 3*d* valence electrons at the interface for the p-AMS can be estimated to be on the order of $10^{14}$ electrons. Furthermore, these additional transferred 3*d* valence electrons from Co are expected to be highly spin-polarized, primarily consisting of spin-down electrons. The magnitude of these delocalized electrons corresponds to a total magnetization on the order of $10^{-6}$ emu, which aligns well with the observed changes in spontaneous magnetization induced by electrical switching, as shown in Figs. 1B, 1C and 2B.

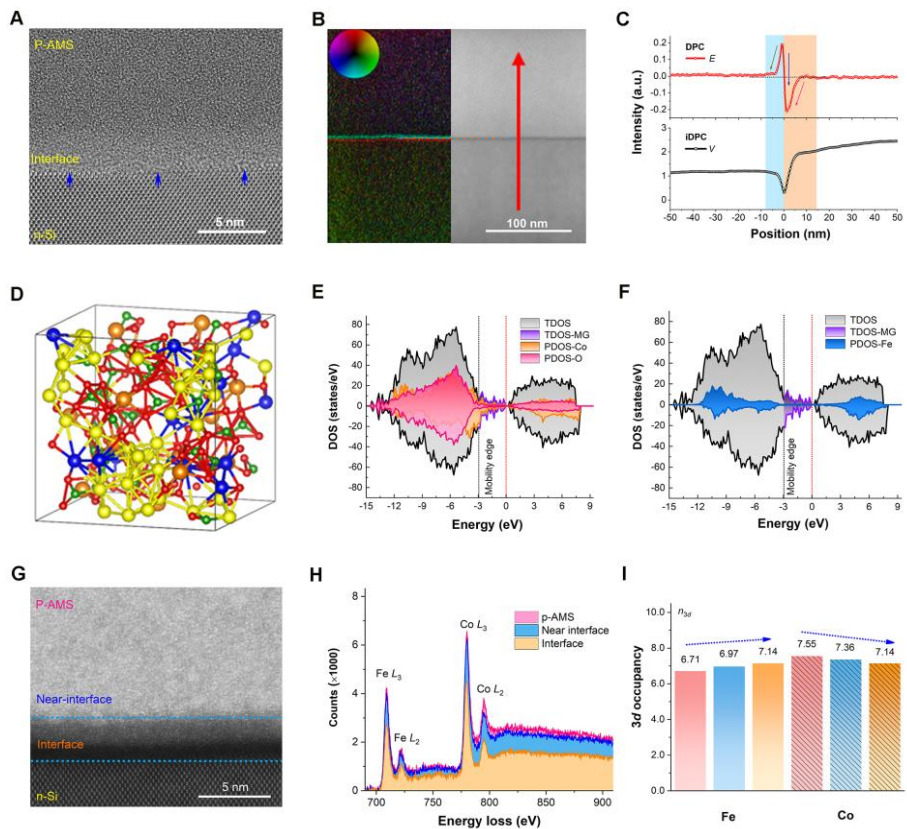

**Fig. 3. Characterization of the magnetic p-n diode.** (**A**) Bright field scanning transmission electron microscopy (BF-STEM) image showing a transition from ordered single-crystalline n-Si to disordered p-AMS. (**B**) Differential phase contrast (DPC) showing the electric-field strength distribution (left side) and integrated differential phase contrast (iDPC) showing 2-dimention projected electrostatic potential (right side). (**C**) Line profile illustrating the changes of electrostatic potential (*V*) and electric-field strength (*E*) with the position indicated in (**B**). (**D**) Atomic model of p-AMS featuring three-dimensional disordered amorphous structure. (**E**) The electronic band structure of p-AMS with a delocalized valence band mainly from Co-3*d* bands with shallow acceptor states in the mobility gap (MG). (**F**) Total density of states (TDOSs) of p-AMS and partial density of states (PDOSs) of Fe indicating Fe-3$d$ bands are localized. (**G**) High angle annular dark field (HAADF) image of the cross-section. (**H**) Electron energy-loss spectra (EELS) for different regions of the magnetic p-n junction diode. (**I**) The calculated average occupation states of the 3$d$ orbitals ($n_{3d}$) for both Co and Fe for different regions

**Discussion**

Through transport, electrostatic potential and electric field strength distributions, and EELS measurements, as well as the first-principles calculations, we found that extended hole states of the p-AMS are spin polarized and the observed spin behavior is attributable to space charge effects. At forward currents of 1 mA and 5 mA (Fig. 4A), the majority hole drift in the p-AMS dominates charge transport, leading to a decrease in spin-polarized hole concentration at the space charge region. This reduction in hole concentration results in a decrease in hole-mediated magnetization. Conversely, at a reverse leakage current of -1 mA, minority electron drift becomes the dominant mechanism for charge transport, facilitating an increase in hole concentration. As a result, hole-mediated magnetization is enhanced. The spin polarization of both majority and minority carriers in the p-AMS is comparable, which accounts for the similar magnetization modulation effects observed at ±1 mA.

At the reverse breakdown current of -5 mA, electron tunneling takes precedence in charge transport as indicated in Fig. 4A. Spin-polarized valence electrons from Co are expected to tunnel into n-Si, promoting reverse accumulation of holes in the p-AMS. These tunneled electrons are almost entirely spin-polarized, which induces the observed giant magnetization amplification in the charge-and-spin diode.

These findings suggest that the space charge region at the magnetic p-n junction functions as a spin modulator by creating distinct junction barriers for different types of charge carriers. This results in direction-dependent selective charge transport processes—including minority drift, electron tunneling, and majority drift—leading to varied spin-based magnetic rectification and amplification effects in the charge-and-spin diode. Furthermore, this spin-modulating effect can be tuned by adjusting the relative positions of the Fermi levels of the integrated materials. Either minority electron drift or tunneling may occur when integrated with n-Si of varying resistivities, while majority hole drift is expected when integrated with p-Si.

Notably, the width of the space charge region at the p-AMS side is approximately 10 nm, as observed in the line profile of the built-in electric field distribution shown in Fig. 3C. When depositing p-AMS layers with thicknesses on the order of this width, the formation of different heterojunction interfaces should significantly influence the overall electrical and ferromagnetic properties.

To confirm this, we deposited p-AMS$_{\sim0.55}$ layers with a thickness of approximately 14 nm onto highly conductive hole-doped p-Si (p1: ~$10^{-3}$ Ωcm), electron-doped n-Si (n1: ~$10^{-3}$ Ωcm), and insulating SiO$_2$ substrates. As shown in Fig. 4B, the *I-U* curves exhibit distinctly different electrical transport behaviors, demonstrating resistance, rectification, and insulation characteristics, respectively. Correspondingly, the p-AMS$_{\sim0.55}$/p1 heterojunction approaches a nearly paramagnetic state due to carrier depletion, as illustrated in Fig. 4C. Remarkably, when integrated with n1, the p-AMS$_{\sim0.55}$ exhibits significantly increased magnetic moments, with a saturation magnetization of about 509 emu/cm³, representing a 29-fold enhancement compared to that of the p-AMS$_{\sim0.55}$/p1 heterojunction. Additionally, the p-AMS$_{\sim0.55}$ in the p-AMS$_{\sim0.55}$/n1 charge-and-spin diode shows a threefold increase in the saturation magnetization compared to that in the p-AMS$_{\sim0.55}$/SiO$_2$ heterojunction (Fig. 4C).

In summary, we have successfully created magnetic p-n junctions for charge-and-spin diodes utilizing room-temperature amorphous magnetic semiconductors. These diodes demonstrate dual functionalities of both charge and spin diodes, realizing switching, rectification, and amplification of both electrical current and magnetization through current alone. Furthermore, a direction-dependent giant spin amplification effect has been identified, primarily attributed to the formation of a magnetic space charge region within this charge-and-spin diode. Without magnetic filed, the ability to electrically switch both electrical current and magnetization in the diode at an ultralow current density of approximately $2.5\times10^{-2}$ A/cm$^2$ marks a significant milestone in the development of silicon-based charge-and-spin diodes. Our current work paves the way for the creation of room-temperature, ultralow-power magnetic semiconductor-based spintronic devices that are compatible with existing complementary metal-oxide-semiconductor fabrication processes, potentially addressing the demands of innovative computing paradigms, including quantum computing and in-memory computing.

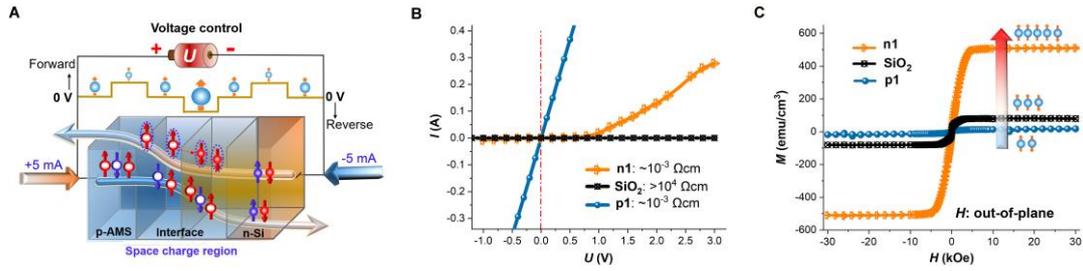

**Fig. 4. Spin modulating effect of p-AMSs deposited on various materials with different $E_F$ levels.** (**A**) Schematic diagrams for the spin modulating effect due to different transport behavior when integrating p-AMSs with n-Si. (**B**–**C**) *I-U* and *M-H* curves of the heterojunctions showing that the integrating material has a substantial effect on regulating the overall electrical and magnetic properties. The heterojunction based on p-AMS can change from a strong ferromagnetic state to a nearly paramagnetic state.

## Methods

### Sample preparation

The p-AMS thin films were deposited on various substrates including single crystalline Si and $SiO_2$ by radio frequency (RF) magnetron sputtering with an alloy target Co-Fe-Ta-B under a gas mixture of pure Ar and $O_2$.

The p-AMS based p-n junction diodes consisting of In/p-AMS (~600 nm)/n-Si (~500 µm)/In were fabricated to probe the electromagnetic responses to temperature ($T$), magnetic field ($H$), and applied bias voltage ($U$). The n-Si substrate has a resistivity of 0.025 ~ 0.026 Ωcm.

### Sample characterization

The thickness of p-AMS thin films was measured by atomic force microscope (AFM, Cypher ES, OXFORD INSTRUMENTS of UK). The atomic structures of the thin film and the p-n junction were observed by a high resolution transmission electron microscopy (HRTEM, JEOL JEM 2010F).

MPMS (Quantum Design) attached with Keithley 2400 source meter and Keithley 6517B electrometer was used for all in-situ magnetic and electrical measurements. During the measurement process, the magnetic field was firstly increased to saturate P-AMS and then decreased to 0 Oe in order to make P-AMS in its remanence state. Subsequently, the transport behavior of the magnetic p-n junction diode was measured at various magnetic fields including 0 Oe, 50 Oe and 100 Oe.

### Electron-energy-loss spectroscopy (EELS) analysis

Scanning transmission electron microscopy (STEM, JEM-ARM300F2AC-TEM) equipped with electron energy-loss spectroscopy (EELS) at an energy resolution of 0.35 eV was employed to analyze the 3$d$ orbital electron state occupancy of ferromagnetic elements Fe and Co. During the experiments, an acceleration voltage of 300 kV and a beam current of 50 pA were utilized, with a pixel time of 0.002 second. DigitalMicrograph software was used to correct the zero-loss peak in the spectrum. Additionally, the Fourier-ratio deconvolution method within the DigitalMicrograph software was applied to eliminate multiple scattering components from the spectrum.

### Differential phase contrast (DPC) imaging

The differential phase contrast (DPC) image was acquired using a Thermo Fisher Scientific Spectra 300 microscope operated at 300 keV and equipped with four segmented detectors. A microprobe mode electron optics configuration was employed, featuring a camera length of 2.3 m and a 70 μm C2 aperture, resulting in a beam convergence angle of 1.6 mrad and a probe size of 0.8 nm. This setup provides an optimal balance between the sensitivity of electric field measurements and spatial resolution. A beam current of 80 pA and an exposure time of 16 μs were utilized to ensure sufficient signal acquisition. The electric field images were subsequently integrated through the integrated DPC (iDPC) procedure to generate the electric potential map.

**Ab initio molecular dynamics calculations phase**

First-principles simulations were performed by a generalized gradient approximation (GGA/GGA+U) with Perdew-Burke-Ernzerhof (PBE) formalism, based on the density functional theory (DFT) as implemented in the Vienna ab initio simulation package (VASP). The supercell model of the Co-Fe-Ta-B-O system, consisting of a total of 200 atoms, was adopted as the initial structure. An automatic k-point mesh of $1 \times 1 \times 1$ containing Gamma was used for geometry optimization. The convergence criterion for the electronic self-consistent calculation was set to $10^{-5}$ eV and the magnitude of force on each atom was required to be below 0.005 eV Å$^{-1}$. An isothermal process was simulated at 300 K with 5000 steps. Electron spins were taken into account throughout the simulation.

**Acknowledgements**

This work was supported by the National Natural Science Funding (Grant Nos. 52571187 and 51922053). This work made use of the resources of the National Center for Electron Microscopy in Beijing and Center for Testing and Analysing of Materials of School of Materials Science and Engineering in Tsinghua University is also appreciated. XM thanks the support by the Fundamental Research Funds for the Central Universities (grant number lzujbky-2025-ytA02). XRW acknowledge the support from the University Development Fund of the Chinese University of Hong Kong, Shenzhen, and the Guangdong Provincial Quantum Strategy Special Project.